\documentclass{PoS}

\title{Hard X-ray Variability of the Brightest Swift/BAT AGN}

\ShortTitle{Hard X-ray Variability of the Brightest Swift/BAT AGN}

\author{\speaker{Claudio Ricci}\\%
       ISDC Data Centre for Astrophysics\\
       Geneva Observatory, University of Geneva \\
       E-mail: \email{claudio.ricci@unige.ch}}

\author{S. Paltani\\
        ISDC Data Centre for Astrophysics \\
        Geneva Observatory, University of Geneva \\}

\author{S. Soldi\\
	Laboratoire AIM - CNRS - CEA/DSM - Universit\'e Paris Diderot (UMR 7158), CEA Saclay,\\
	DSM/IRFU/SAp, F91191 Gif-sur-Yvette, France\\}
	
\author{T. J.-L. Courvoisier\\
        ISDC Data Centre for Astrophysics \\
        Geneva Observatory, University of Geneva \\}

\abstract{ Variability is one of the hallmarks of Active Galactic Nuclei. The Burst Alert Telescope onboard of {\it Swift}, with its homogeneous coverage of the sky is a formidable tool to study variability at hard X-rays. We present here the analysis of the 1-month binned {\it Swift}/BAT lightcurves of the 20 brightest Active Galactic Nuclei in the hard X-ray sky. The sample consists of 2 blazars, 3 radio galaxies, 6 Seyfert 1/1.5s, 8 Seyfert 2s and 1 Narrow Line Seyfert 1. We found that all the objects show variability, and most of them have a value of the fractional root mean squared variability amplitude of $F_{\rm\,var}\sim 0.2-0.3$. We did not find any significant correlation of $F_{\rm\,var}$ with the column density or the luminosity in our sample.}

\FullConference{Fast X-ray timing and spectroscopy at extreme count rates\\
		 February 7-11, 2011\\
		Champ\'ery, Switzerland}

\begin{document}

\section{Introduction}
Active Galactic Nuclei (AGN) are amongst the most luminous X-ray sources in the sky. AGN are thought to be powered by accretion onto supermassive black holes (Rees, 1984), with their X-ray emission probably originating in a hot corona sandwiching the accretion disk (Haardt \& Maraschi, 1991) in radio-quiet objects, and in the jet in radio-loud AGN (e.g., Boettcher 2010). Variability is one of the key features of AGN, and it was found to be significative in the X-ray band already in early observations of nearby Seyfert galaxies (Sy) performed by {\it Ariel V} (Marshall et al., 1981). The X-ray variability of AGN is aperiodic, and their power spectral density distribution (PSD) can be normally described with a broken power law, with indices ranging between $-1$ and $-2$ (McHardy \& Czerny, 1987).

The Burst Alert Telescope (BAT) onboard of {\it Swift} (Barthelmy et al., 2005) scans continuously the whole sky in the 14--195\,keV energy range, and is thus an extremely well suited instrument for studying AGN variability at hard X-rays. Here we report a study of the hard X-ray variability of a small sample of AGN. The sample consists of the 20 brightest AGN detected by {\it Swift}/BAT, of these 2 are blazars, 2 narrow-line radio galaxies (NLRG), 1 broad-line radio galaxy (BLRG), 6 Seyfert\,1/1.5s, 8 Seyfert\,2s and 1 Narrow Line Seyfert\,1 (NLS1). The 1-month binned light curves have been taken from the NASA {\it Swift}/BAT 58 months catalog\footnote{http://swift.gsfc.nasa.gov/docs/swift/results/bs58mon/} (Baumgartner et al., 2011).

\section{Variability estimators}
A way to estimate the variability is through the fractional root mean squared (rms) variability amplitude $F_{\rm\,var}$ (Edelson et al., 1990), defined as
\begin{equation}
F_{\rm\,var}= \sqrt{\frac{S^2-\overline{\sigma_{err}^2}}{\bar{x}^2}}.
\end{equation}
Where the sample variance $S^2$ is given by
\begin{equation}
S^2=\frac{1}{N-1}\sum_{i=1}^{N}(x_i -\overline{x})^2,
\end{equation}
while the mean square error $\overline{\sigma_{err}^2}$ by 
\begin{equation}
\overline{\sigma_{err}^2}=\frac{1}{N}\sum_{i=1}^{N}\sigma^2_{err,i}.
\end{equation}
The error of $F_{\rm\,var}$ is given by 
\begin{equation}
err(F_{\rm\,var})=\sqrt{\left(\sqrt{\frac{1}{2N}}\cdot  \frac{\overline{\sigma_{err}^2}}{\overline{x}^2F_{\rm\,var}} \right)^2+\left(\sqrt{\frac{\overline{\sigma_{err}^2}}{N}}\cdot\frac{1}{\overline{x}}\right)^2}.
\end{equation}
In the following we will use $F_{\rm\,var}$ to characterize the variability of the objects in our sample.

\begin{table*}[t]
\centering
\caption{Properties of the sources of our sample: (1) detection significances, (2) luminosities in the 14--195\,keV energy range, (3) fractional rms variability amplitudes on a timescale of 30 days,  (4) hydrogen column densities. }
\smallskip
\label{tab:fvar}
\begin{tabular}[c]{lccccc}
\hline \hline \noalign{\smallskip}
	&   (1)       & (2) & (3) & (4) &  \\

 Source  	&     Det. Significance       & $\log L_{14-195\rm \,keV}$ &     $\rm\,F_{\,var}$ {\scriptsize (30-days)} & $N_{\rm\,H}$& Type\\
  	&     {\scriptsize $[\sigma]$}       & {\scriptsize [$\rm\,erg\,s^{-1}$]} & &{\scriptsize [$\rm\,cm^{-2}$]}  \\
\noalign{\smallskip\hrule\smallskip}
%
Cen A   & $ 428.7 $ & $ 44.01  $ & $0.399 \pm 0.002$ & 12$^a$ &      NLRG  \\
NGC 4151   & $ 275.0 $  & $  44.11 $ &  $0.280\pm0.004$ & 6.9$^b$ &      Sy 1.5 \\
3C 273   & $ 156.8 $  & $ 47.47  $ & $0.31 \pm 0.01$  & 0.5$^b$ &      Blazar \\
NGC 4388   & $ 110.7 $ & $ 44.64  $ & $0.31\pm 0.01$  & 27$^b$ &      Sy 2 \\
Mrk 421   & $ 109.5 $ & $ 45.46 $ &  $0.96 \pm 0.01$  & 0.1$^b$ &      Blazar \\
Circinus Galaxy   & $ 101.7 $ & $43.09$ & $0.13\pm0.01$ & 360$^b$&      Sy 2  \\
IC 4329A   & $ 101.1  $  & $  45.22 $   & $0.19\pm 0.02$ & 0.4$^b$ &      Sy 1  \\
NGC 2110 & $98.1$ & 44.60 & $0.32\pm 0.01$  & $4.3^c$ &       Sy2 \\
NGC 5506 & $95.0$ & 44.31 & $0.27\pm 0.02$ & $3.4^c$ &       NLS1 \\
MCG$-$05$-$23$-$016& $90.4$ & 44.50 & $0.21\pm 0.01$   & $1.6^c$&       Sy2 \\
IGR J21247+5058   & $ 83.3 $ & $  45.25 $  & $0.31\pm 0.01$   & $0.6^b$ &      BLRG \\
NGC 4945   & $  76.1 $ & $ 43.37 $ & $0.35 \pm 0.02$  & $400^b$ &       Sy2 \\
Mrk 348  & $  70.4 $ & $ 44.91 $ & $0.28\pm 0.02 $  & $30^c$ &       Sy2\\
NGC 3783  & $  68.7 $ & $ 44.60 $ & $ 0.26\pm 0.02 $  & $0.1^c$ &       Sy1.5 \\
NGC 4507  & $  64.6 $ & $ 44.77 $ & $ 0.29 \pm 0.02 $ & $29^c$  &      Sy 2   \\
NGC 3516   & $ 62.3   $ & $44.33   $ & $  0.30 \pm 0.02   $ & $4^c$  &      Sy 1.5 \\
NGC 7172   & $  60.1  $ & $ 44.46  $ & $ 0.35  \pm 0.04^*   $ & $9^c$ &      Sy 2  \\
NGC 3227   & $ 56.2   $ & $ 43.57 $ & $  0.16 \pm 0.21   $ & $6.8^c$  &      Sy 1.5 \\
Cyg A   & $ 54.0 $ & $ 46.01  $ &   $0.34\pm0.02$ & $11^b$ &      NLRG \\
MCG +08$-$11$-$011   & $ 49.0   $ & $ 45.09  $ & $  0.35 \pm 0.03  $ & $0.2^c$ &       Sy 1.5  \\
\noalign{\smallskip}
{\it Crab}   & $ 7496  $ & -- & $  0.0215 \pm  0.0004 $ & -- & --  \\

\noalign{\smallskip}
\hline

\multicolumn{6}{l}{{\bf Notes.} {\footnotesize $^a$ Beckman et al. (2011), $^b$ Beckmann et al. (2009) and references therein,}} \\
\multicolumn{6}{l}{{\footnotesize $^c$ Ricci et al. (2011) and ref. therein. $^*$ "Flare" of January 2008 removed.}} \\

\end{tabular}
\end{table*}

\newpage
\begin{figure*}[h!]
\centering
\begin{minipage}[!b]{.48\textwidth}
\centering
\includegraphics[width=7.5cm]{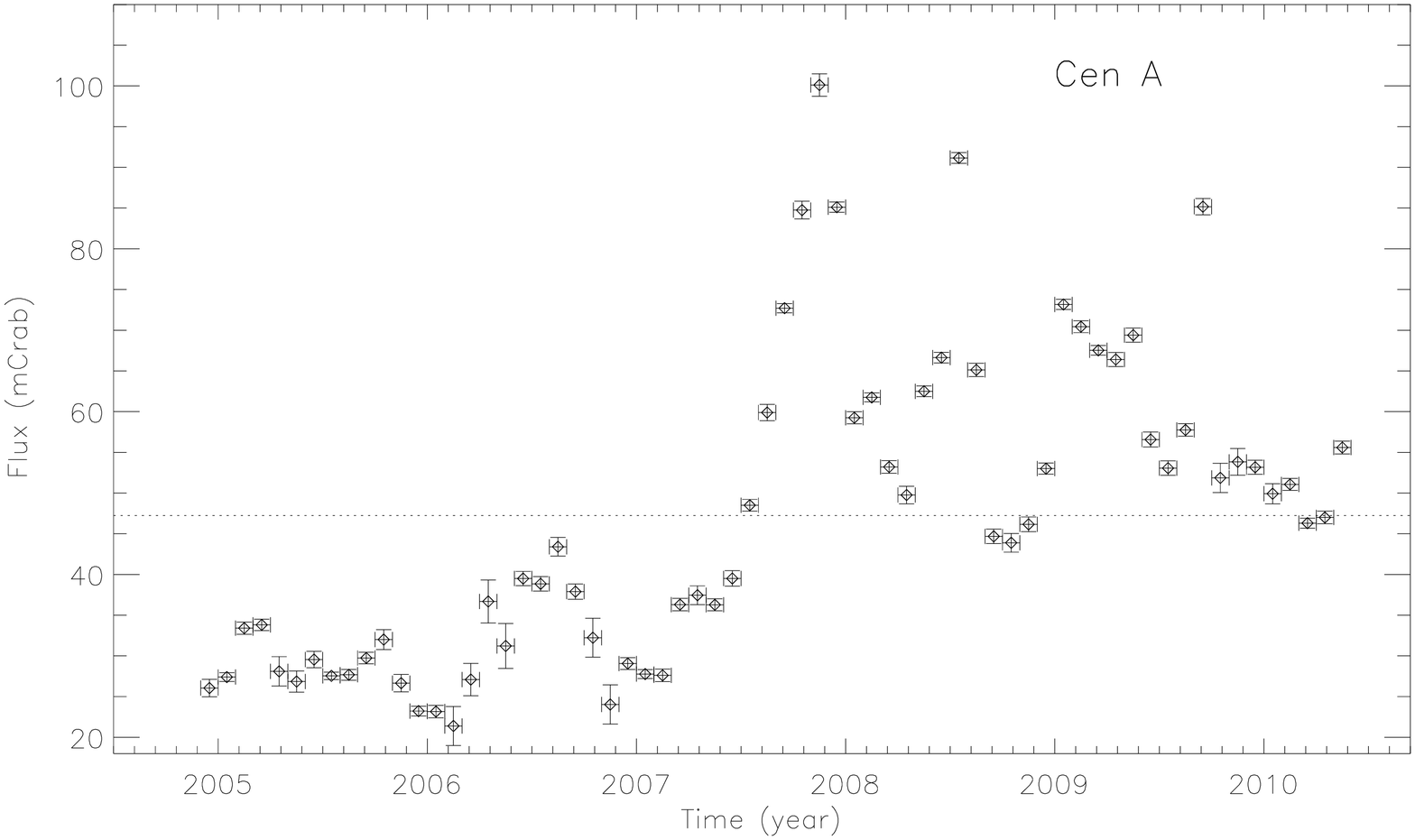}
\end{minipage}
\hspace{0.05cm}
\begin{minipage}[!b]{.48\textwidth}
\centering
\includegraphics[width=7.5cm]{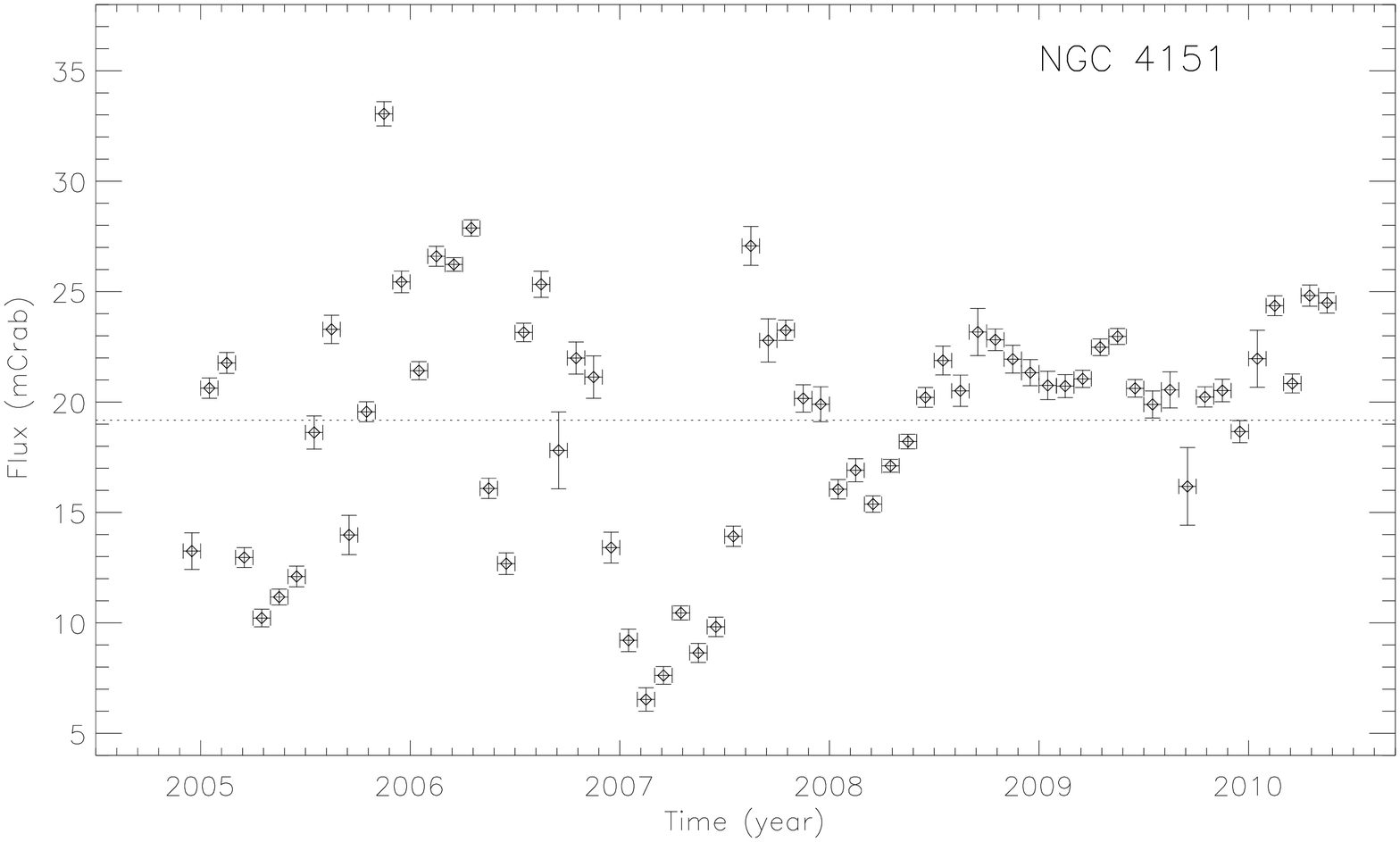} \end{minipage}
 \begin{minipage}[!b]{.48\textwidth}
\centering
\includegraphics[width=7.5cm]{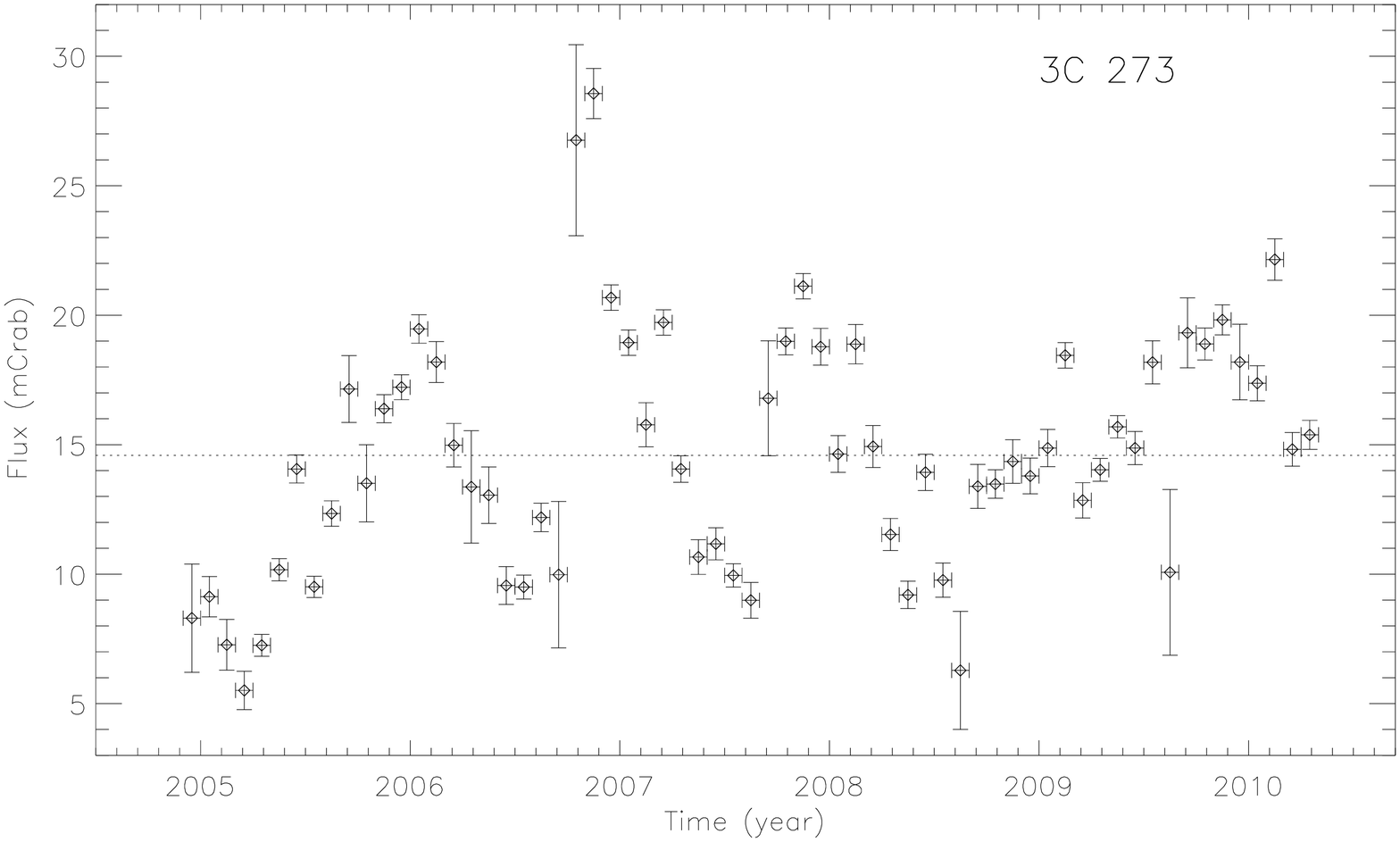}
\end{minipage}
\hspace{0.05cm}
\begin{minipage}[!b]{.48\textwidth}
\centering
\includegraphics[width=7.5cm]{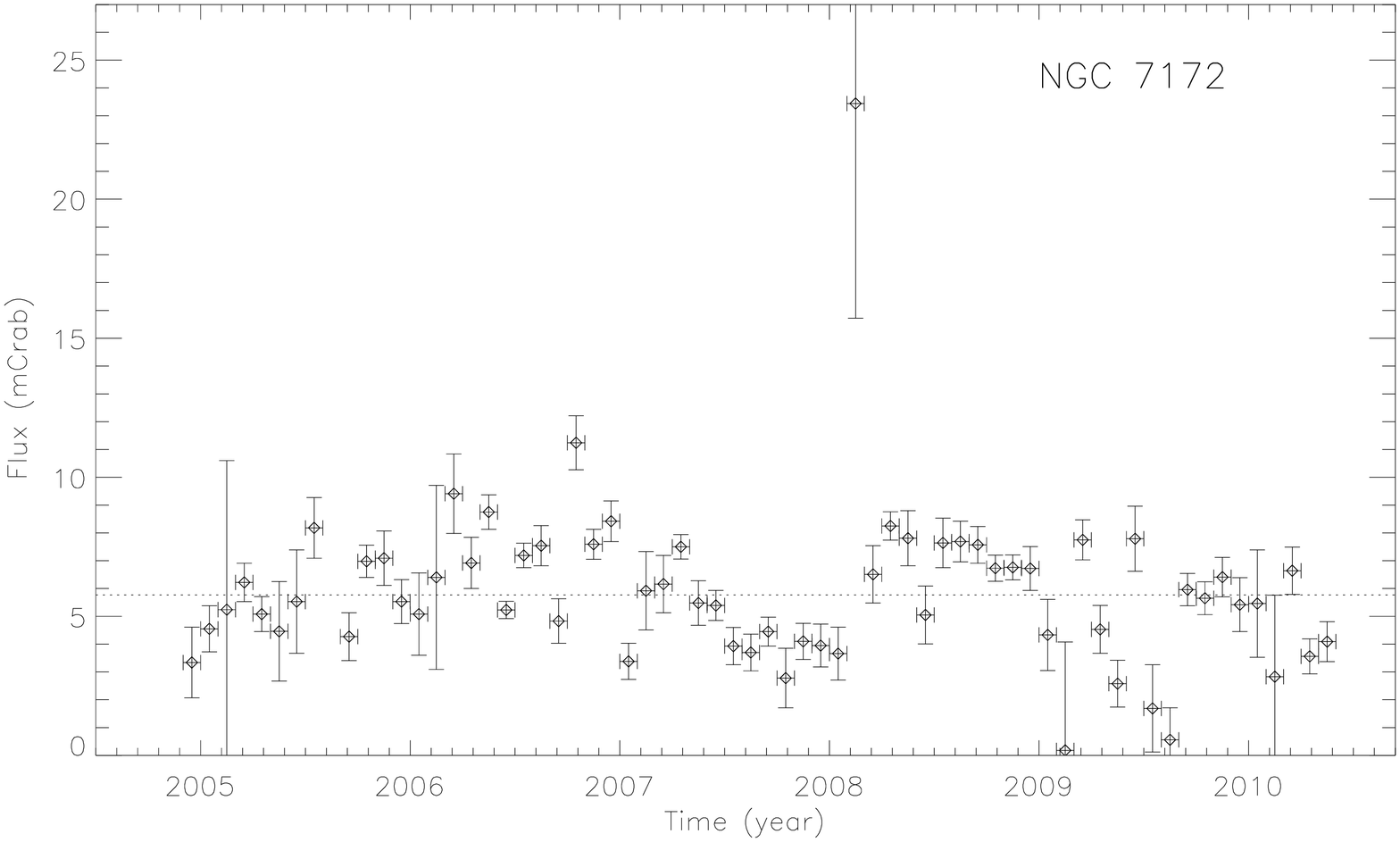}\end{minipage}

\begin{minipage}[!b]{.48\textwidth}
\centering
\includegraphics[width=7.5cm]{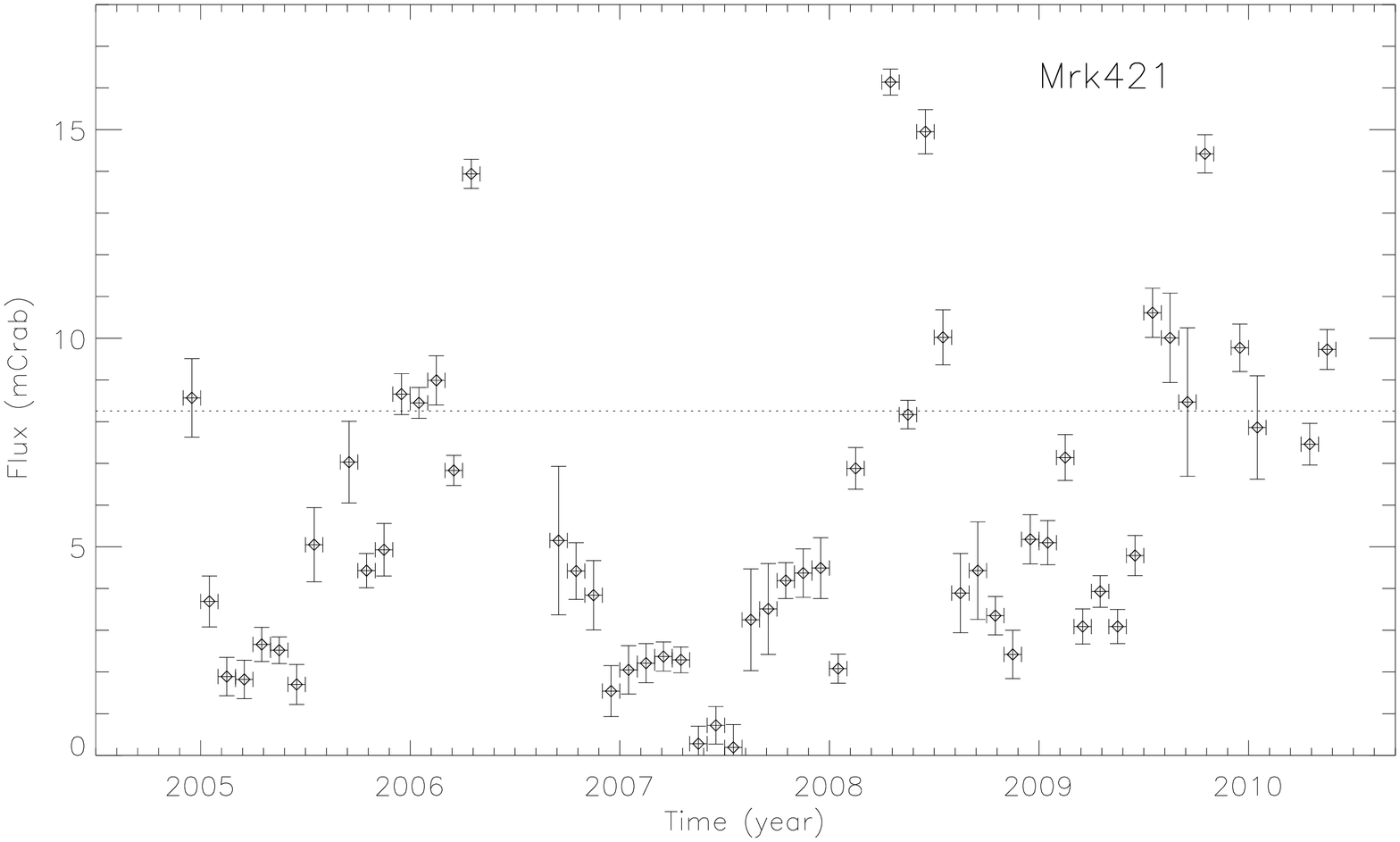}
\end{minipage}
\hspace{0.05cm}
\begin{minipage}[!b]{.48\textwidth}
\centering
\includegraphics[width=7.5cm]{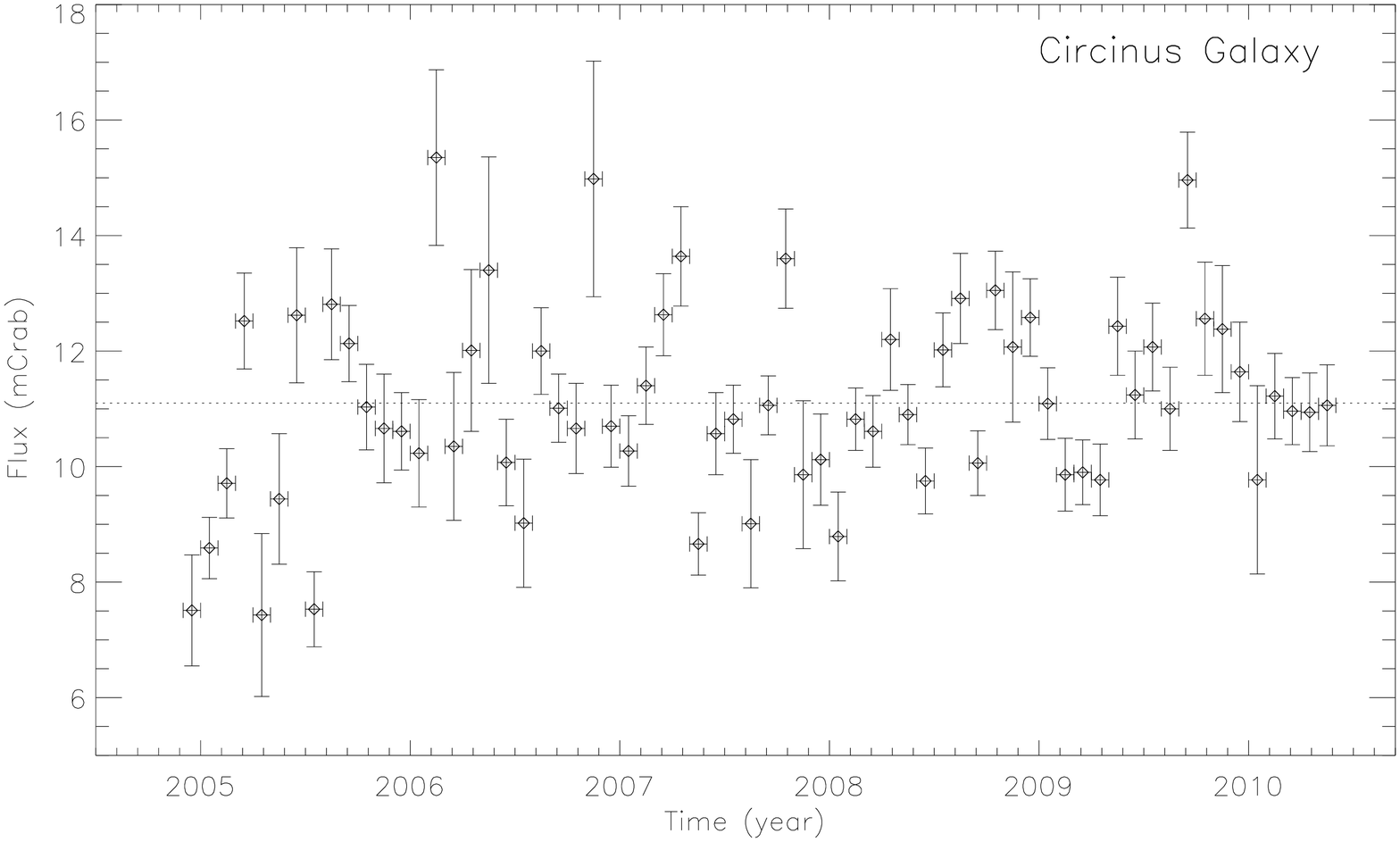} \end{minipage}
 \begin{minipage}[t]{1\textwidth}
  \caption{{\it Swift/BAT} 30-days binned light curves in the 14--195\,keV band. The dotted horizontal lines represent the average value for each object.}
\label{fig:agn_lc}
 \end{minipage}

\end{figure*}

\begin{figure*}[h!]
\centering
\includegraphics[width=10cm]{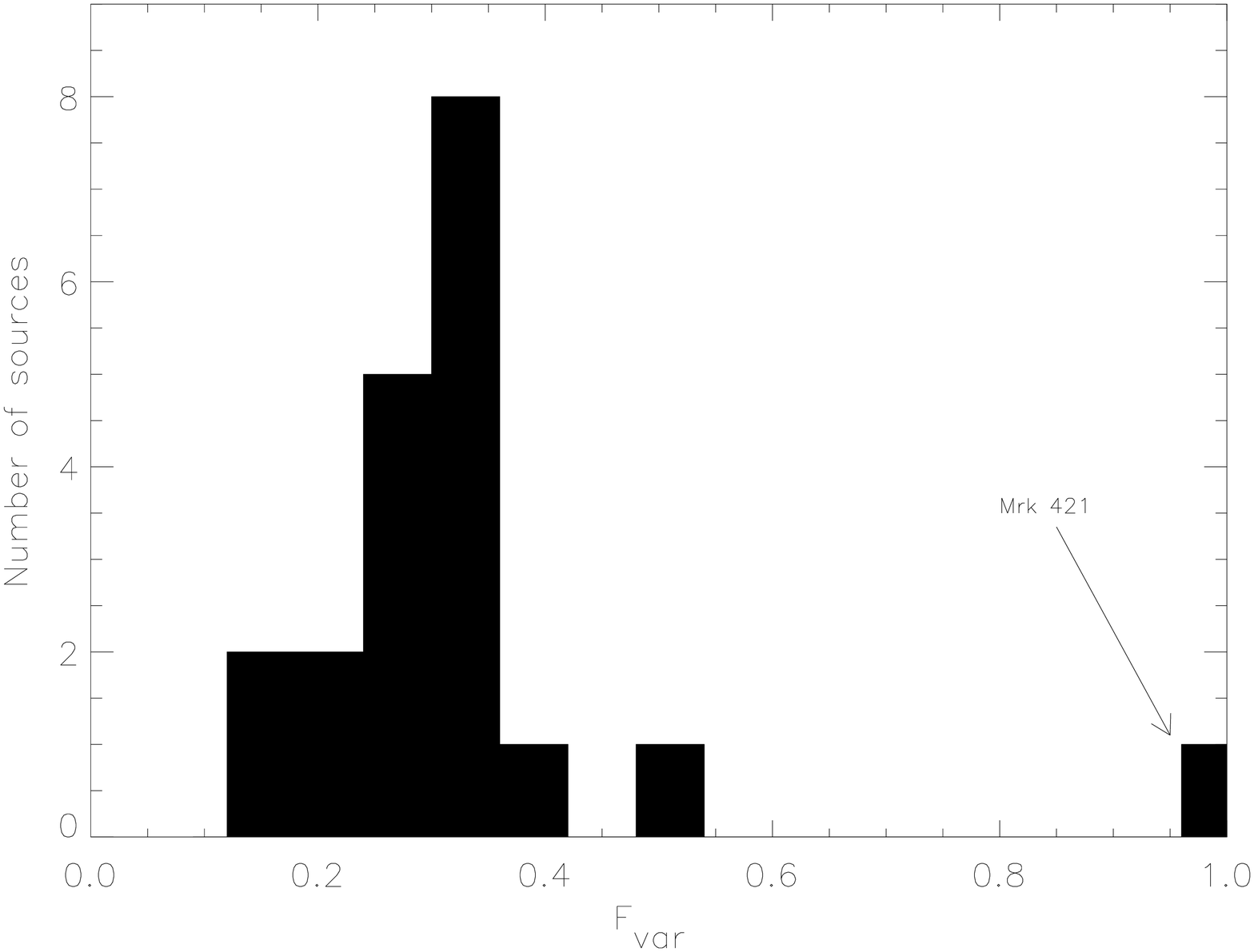}
\caption{Distribution of the fractional rms variability amplitude for the 20 sources of our sample.}
\label{fig:fvardistr}
\end{figure*}

\section{Results}

In Table\,\ref{tab:fvar} are listed the values of the fractional rms variability amplitude of the objects of our sample, and as a check, the value obtained from the light curve of the Crab ($F_{\rm\,var}\sim0.02$). The value of the Crab can be associated to the systematic error of the {\it Swift}/BAT data. In Fig.\,\ref{fig:agn_lc}, we show the light curves of 6 out of the 20 sources of our sample. All the sources of our sample show hard X-ray variability on the time-scale of one month.
As it can also be seen from Fig.\,\ref{fig:fvardistr}, the value of the fractional rms variability amplitude is $F_{\rm\,var}\sim0.2-0.3$ for most of the objects of the sample, with the average value being $\overline{F_{\rm\,var}}=0.32$. The blazar Mrk\,421 shows a much stronger variability ($F_{\rm\,var}\sim 0.96$) than the average value of the sample. Amongst the radio-quiet NGC\,7172 is the most variable, with a value of $F_{\rm\,var}\sim 0.48$. This is due to what would appear to be a flare, registered in January 2008. Excluding this outlier point NGC\,7172 shows a variability consistent with the average of our sample ($F_{\rm\,var}=0.35\pm0.04$).
At the other end of the distribution, the Compton-thick Seyfert\,2 Circinus Galaxy and the Seyfert\,1 IC\,4329A show the smallest amounts of variability ($F_{\rm\,var}\sim 0.13$ and $F_{\rm\,var}\sim 0.19$, respectively). The low value of $F_{\rm\,var}$ of Circinus Galaxy is very likely related to the reflection-dominated nature of its hard X-ray spectrum.

\begin{figure*}[h!]
\centering
\includegraphics[width=10cm]{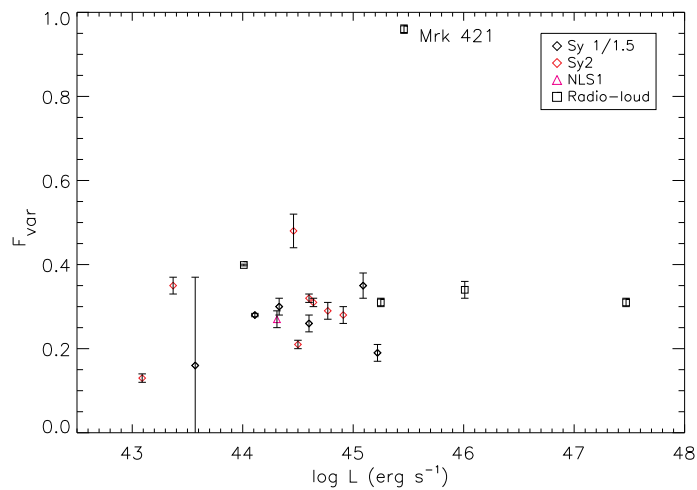}
\caption{$F_{\rm\,var}$ versus luminosity for the sources of our sample.}
\label{fig:corr1}
\end{figure*}

\begin{figure*}[h!]
\centering
\includegraphics[width=10cm]{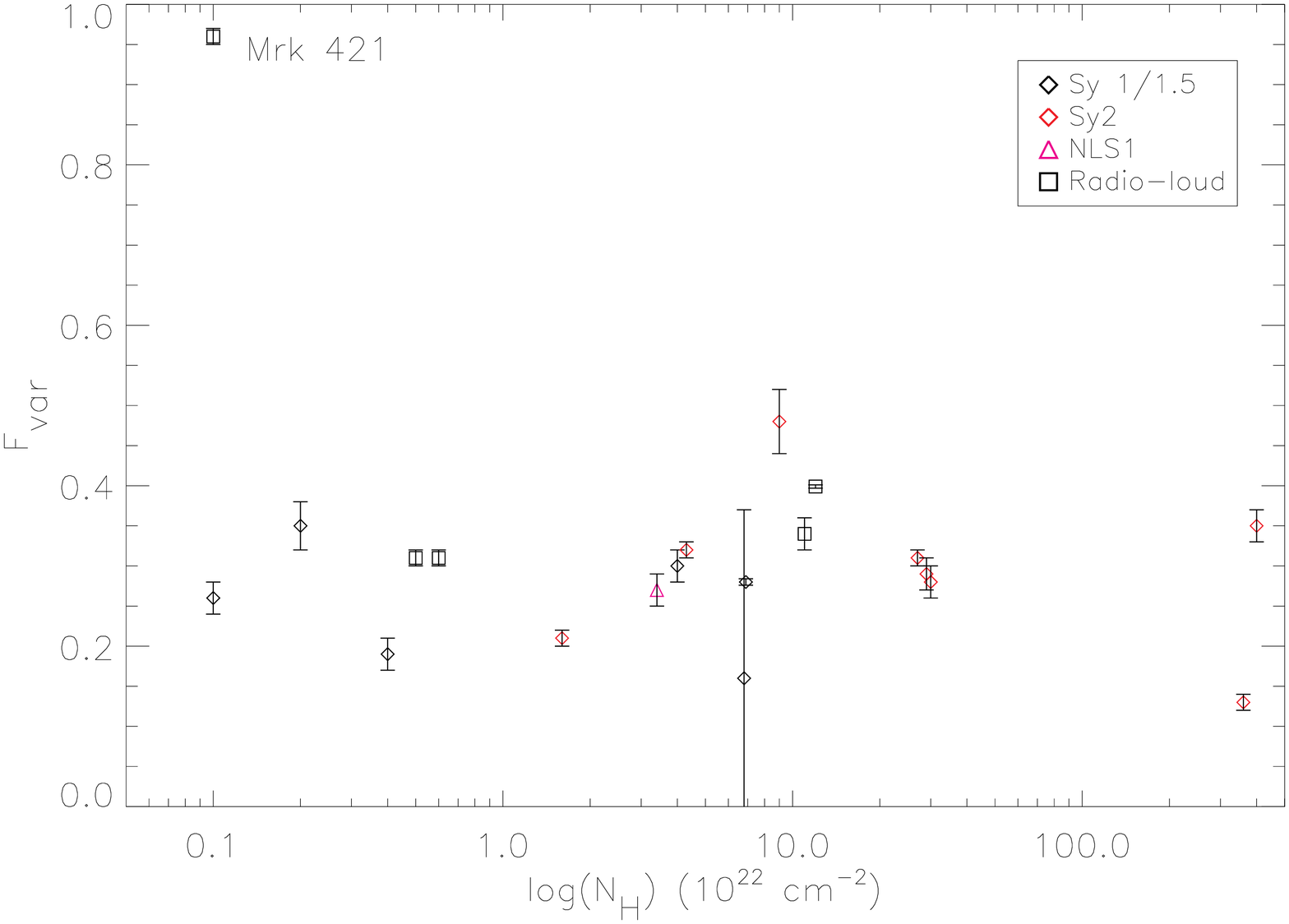}
\caption{$F_{\rm\,var}$ versus hydrogen column density for the sources of our sample.}
\label{fig:corr2}
\end{figure*}

\section{Variability vs Luminosity and Column density}
An inverse correlation between the variability amplitude in the X-rays and the X-ray luminosity of AGN was found by (Barr et al., 1986)using {\it EXOSAT} data. More recently, Beckmann et al. (2007) studied the hard X-ray variability of the 44 brightest AGN detected by BAT after 9 months of operations, and found that possibly this anti-correlation is extended to the hard X-ray band (see also Soldi et al., 2010). They also found a possible correlation between the hydrogen column density $N_{\rm\,H}$ and the variability amplitude. 
We investigated the existence of these two correlations in our sample (see Figs.\,\ref{fig:corr1} and \ref{fig:corr2}). The variability amplitudes in our sample are confined in a small range of values, and no correlation with other parameters is evident. A Spearman rank test gives a correlation coefficient between $F_{\rm\,var}$ and the luminosity of $r_s=0.27$ , while it is $r_s=0$ between $F_{\rm\,var}$ and $N_{\rm\,H}$. These values correspond to a probability of correlation of $78\%$ in the first case, and of $0\%$ in the second case. Similar results are obtained also considering Mrk\,421 as an outlier.  No significant correlation is found also dividing the sample in three categories (Sy 1/1.5, Sy2, radio-loud). The lack of correlations might be due to the limited sample we used, and further studies are needed to better probe it.

\section{Conclusions}
Studying the 1-month binned {\it Swift}/BAT light-curves of the 20 brightest objects after 58 months of observations, we found that all the objects in our sample show variability, ranging between $F_{\rm\,var}=0.13$ and $F_{\rm\,var}=0.96$, for Circinus galaxy and Mrk\,421, respectively. The average value of the variability amplitude if $F_{\rm\,var}\sim0.3$. We did not find any significant correlation of the variability amplitude with the luminosity and the hydrogen column density.

\end{document}